\documentclass[a4paper,12pt]{article}
\usepackage[utf8]{inputenc}
\usepackage[T1]{fontenc}
\usepackage[english]{babel}
\usepackage{color}
\usepackage{amssymb}
\usepackage{amsmath}
\usepackage{ae}
\usepackage{slashed}
\usepackage{graphics}
\usepackage{graphicx}
\usepackage{epstopdf}
\usepackage{url}

\def\rmi{{\rm i}}

\newcommand{\g}{\gamma}

\newcommand{\jb}{\bar{\jmath}}

\newcommand{\lp}{\left(}

\newcommand{\rp}{\right)}
\newcommand{\BS}{\bf S}
\newcommand{\BC}{\bf C}
\newcommand{\BSbar}{\bf \bar{S}}

\newcommand{\BPhi}{\bf \Phi}
\newcommand{\BPhibar}{\bf \bar \Phi}
\newcommand{\BB}{\mathbf{B}}
\newcommand{\K}{K\"ahler}
\newcommand{\be}{\begin{equation}}
\newcommand{\ee}{\end{equation}}
\newcommand{\bea}{\begin{align}}
\newcommand{\eea}{\end{align}}
\newcommand{\li}{\hspace{1mm}}

\newcommand{\mK}{\mathcal{K}}
\usepackage{cite}
\usepackage{mciteplus}
\usepackage{epsfig}
\usepackage{times}
\usepackage[small]{caption}
\usepackage{ifthen}
\usepackage{longtable}
\usepackage{multirow}
\newcommand{\rf}[1]{(\ref{#1})}
\newcommand{\bbox}{\lower.2ex\hbox{$\Box$}}

\usepackage[colorlinks=true,linktoc=page,linkcolor=blue,citecolor=blue,filecolor=blue,urlcolor=blue]{hyperref}
\usepackage{pstricks}

\usepackage{color}
\definecolor{darkgreen}{rgb}{0,.5,0}

\newcommand{\vp}{\varphi}

\renewcommand{\O}{\Omega}
\newcommand{\C}{\mathcal{C}}
\newcommand{\Z}{\mathcal{Z}}
\renewcommand{\H}{\mathcal{H}}
\newcommand{\B}{\mathcal{B}_\mu}
\renewcommand{\L}{\Lambda}
\newcommand{\D}{\mathcal{D}}

\newcommand{\Op}{\Omega^\phi}
\def\rmi{{\rm i}}
\newcommand{\ba}{\begin{eqnarray}}
\newcommand{\ea}{\end{eqnarray}}
\numberwithin{equation}{section}

\begin{document}
\selectlanguage{english}
\setcounter{secnumdepth}{3}
\frenchspacing
\pagenumbering{roman}

\null{\vspace{\stretch{1}}}

\begin{center}
{\Large {\bf Orthogonal Nilpotent Superfields\\

\

from Linear Models}}\\
\vspace{\stretch{1}}
{\normalsize Renata Kallosh, Anna Karlsson, Benjamin Mosk and Divyanshu Murli}\\
\vspace{\stretch{1}}
{\small Stanford Institute for Theoretical Physics and Department of Physics,\\ Stanford University, Stanford, CA 94305 USA}\\
\vspace{\stretch{1}}
\end{center}

\begin{abstract}\noindent
We derive supersymmetry/supergravity models with constrained orthogonal nilpotent superfields from the linear models in the formal limit where the masses of the sgoldstino, inflatino and sinflaton tend to infinity. The case where the sinflaton mass remains finite leads to a model with a `relaxed' constraint, where the sinflaton remains an independent field. Our procedure is equivalent to a requirement that some of the components of the curvature of the moduli space tend to infinity.
 
\end{abstract}
\vspace{\stretch{2}}
\noindent\makebox[\linewidth]{\rule{\textwidth}{0.4pt}}
{\footnotesize {e-mails: kallosh@stanford.edu, annakarl@stanford.edu, bmosk1@stanford.edu, divyansh@stanford.edu}}
\thispagestyle{empty}
\newpage
\pagenumbering{arabic}
\tableofcontents{}
\section{Introduction}
The supergravity models with orthogonal nilpotent superfields appear to be very useful in cosmology \cite{Ferrara:2015tyn,Carrasco:2015iij}, for example with respect to supersymmetry being realized non-linearly. The purpose of this paper is to find out if the absence of the inflatino and the sinflaton in the spectrum of the non-linear models can also be understood by requiring the existence of the formal limit of masses of the corresponding particles going to infinity in the linear supergravity models. 

An analogous situation has been studied in \cite{Kallosh:2015pho}, where the linear model in the limit of the infinite mass of the sgoldstino was shown to lead to a theory with a nilpotent multiplet 
\be \label{eq.nilpS}
{\BS^2}=0.
\ee 
This theory has non-linearly realized supersymmetry and contains no fundamental sgoldstino, in line with the fact that the corresponding Volkov-Akulov (VA) model \cite{Volkov:1972jx, *Volkov:1973ix} has spontaneously broken global supersymmetry and only a fermion field, i.e. a spectrum without bosons. The relation between linear and non-linear supersymmetry was investigated in \cite{Rocek:1978nb, *Ivanov:1978mx, *Lindstrom:1979kq, *Samuel:1982uh, *Casalbuoni:1988xh}, where the nilpotent chiral multiplet ${\bf S}^2=0$ was proposed with regard to the VA theory. Other constraints on superfields in global supersymmetry were studied in \cite{Komargodski:2009rz,Dudas:2011kt}, and the theory with a constrained nilpotent superfield ${\bf S}^2=0$ was shown to be equivalent to a VA model in \cite{Kuzenko:2010ef}.
 Bosonic supergravity models of the VA type, with constrained superfields in application to inflation, were proposed in 
\cite{Antoniadis:2014oya, *Dudas:2015eha}.

Another way to describe the linear model underlying the model with non-linearly realized supersymmetry in the case of the nilpotent multiplet ${\BS^2}=0$ is to introduce the Lagrange multiplier superfield ${\bf \Lambda}$ to the constraint of the form ${\bf \Lambda} {\BS^2}$. This has been done both in the global supersymmetry model \cite{Kuzenko:2011tj} and in the local superconformal theory \cite{Ferrara:2014kva}.

The orthogonal nilpotent superfields studied in \cite{Komargodski:2009rz,Kahn:2015mla,Ferrara:2015tyn,Carrasco:2015iij,Dall'Agata:2015lek} are the chiral field $\BS$, nilpotent by \rf{eq.nilpS}, and a real superfield
\be \label{eq.B}
 \BB=\frac{1}{2i}(\BPhi-\BPhibar),
\ee
related by the orthogonality condition 
\be \label{orth}
{\BS \BB}=0.
\ee
This sets the components of the superfield $\BPhi$, an inflaton superfield in application to cosmology, to be functionals of the $\BS$ multiplet, rather than fundamental. A relaxed constraint \cite{Komargodski:2009rz,Dudas:2011kt,Kahn:2015mla,Ferrara:2015tyn,Carrasco:2015iij,Dall'Agata:2015lek} 
\be
\bar D_{\dot \alpha}({\BS{\bf B}})=0
\label{relaxed} \ee
allows the scalar component of $\BPhi$ to remain independent.

In the setting of Lagrange multipliers, a complete description of these constraints remains to be formulated, although it has been proposed \cite{Ferrara:2015tyn} that such a Lagrange multiplier has to be a complex general superfield. A detailed investigation of general Lagrange multipliers for various constrained superfields is performed in \cite{FKVW}, on the basis of the supermultiplet tensor calculus for ${\cal N}=1$ supergravity \cite{Kugo:1983mv}, adapted to the notation of \cite{Freedman:2012zz}. A review of the supermultiplet calculus and many references can be found in \cite{FKVW}.

Here we provide an alternative version of the linear supergravity model, without Lagrange multipliers, by analysing the limits of models with linear supersymmetry when the masses of the relevant particles tend to infinity. In these limits we show how the models with constrained orthogonal nilpotent superfields \rf{orth}, or the relaxed constraint \rf{relaxed}, emerge. A previous attempt \cite{Dudas:2011kt} to the latter was unsuccessful: a discrepancy between the superfield constraints and the limit to heavy fermions in $\BPhi$ was revealed. However, the choice of the `microscopic theory' in \cite{Dudas:2011kt} was not suitable for this purpose, as we show in appendix \ref{app.EarlierWork}.

In the global supersymmetry case the procedure is relatively simple. Terms with curvature of the moduli space, related to the masses of some fields, are added to the \K\, potential. Such models are shown to have a well-defined limit when the masses tend to infinity: the limiting models are those with constrained superfields. We argue that the constraints are sufficient and necessary conditions for the existence of such a limit.

In the local case we use the general multiplet calculus presented in \cite{FKVW}, and we apply it to our models which in the infinite mass limit lead to orthogonal nilpotent constraints, or relaxed constraints, in supergravity. This result is closely related to the result presented in \cite{Ferrara:2015tyn}, but we make one more step and produce the solution of the constraints in the case of local supersymmetry.

\section{Global supersymmetry and the moduli space curvature}
Consider the model with two standard chiral superfields, where\footnote{Note that we will use the notation $\chi=P_L\chi$ and $\bar{\chi}=P_R\chi$, with $(\chi)^2=\chi P_L\chi$ and $(\bar{\chi})^2=\chi P_R \chi$, and follow the conventions of \cite{notationCam}, in the 2-component setting.}
\be
{\BS}= s(x) + \sqrt 2 \theta \chi^s + \theta^2 F^s
\label{S}\ee
is often called the stabilizer superfield, where $\chi^s$ is the goldstino, $s$ is the sgoldstino and the auxiliary field $F^s$ is the order parameter for supersymmetry breaking. The inflaton superfield is given by
\be\label{eq.PhiField}
{\BPhi}= \varphi+i b(x) + \sqrt 2 \theta \chi^\phi + \theta^2 F^\phi,
\ee
where $\varphi$ is the inflaton, $b$ is the sinflaton and $\chi^{\phi}$ is the inflatino. Global supersymmetry is linearly realized and the initial flat \K\, potential is of the form 
\be
K_{\rm flat}(\BS, \BSbar; \BPhi , \BPhibar ) = \BS \BSbar + \BB^2,
\label{flat}\ee
with the real superfield $\BB$ given by \rf{eq.B}. This \K\, potential is invariant under shifts of the inflaton field $\varphi$. The superpotential is a general holomorphic function of both superfields: 
\be
W= W(\BS, \BPhi).
\ee

To relate this linearly supersymmetric model to a model with non-linearly realized supersymmetry, we add the following terms to the \K\, potential:
\be \label{Delta}
\Delta K= - c {(\BS \BSbar)^2} + c_1 {\BS}\BB^2 + \bar c_1 {\BSbar}\BB^2 - { c_2\, \BS\BSbar\BB^2}.
\ee
Here, the four arbitrary constants are $c$, $c_1=a_1+i b_1$ and $c_2$, all independent of each other. In supergravity, as long as all $c_i$ are finite and independent, we have a perfectly consistent linearly realized supergravity model, but the moduli space is not flat. As we will see below, these constants are defining various components of the moduli space curvature. The sectional curvature ${(\BS \BSbar)^2}$ and the bisectional curvature $\BS\BSbar\BB^2$ were proposed in \cite{Kallosh:2010xz} with the purpose to generate the masses of the sgolsdstino and the sinflaton, respectively. The cubic term ${\BS}\BB^2$ was suggested in \cite{Ferrara:2015tyn} with the purpose to generate the mass of the inflatino.

In general, the moduli space curvature is defined as
\be
R_{i \jb k \bar l}= {\cal K}_{i \jb k \bar l} - \Gamma^m_{ik} g_{m\bar m} \Gamma^{ \bar m}_{\jb\bar l}\, , \quad R_{i j k \bar l}= {\cal K}_{i j k \bar l}\li.
\label{curv}\ee
The curvature terms affect the masses of the bosonic particles as follows \cite{GomezReino:2006dk}
\be
\Delta{\cal M}_{i\jb}= -R_{i\jb k \bar l} \bar F^k F^{\bar l} 
\label{mass}\ee
where $\bar F^k, F^{\bar l}$ are auxiliary fields, assuming that the Christoffel symbols vanish. Here we are using the framework and notation as defined in \cite{Kallosh:2010xz}, and we will only look at constant, field-independent contributions, evaluating the mass terms at $\BS=\BSbar=\BB=0$. The fermionic masses are set by:
\be
\Delta m_{ij} = - \Gamma_{ij}^k F_{k}\, , 
\label{mass1}\ee
following \cite{Freedman:2012zz}. 

The new feature of our models described by \rf{Delta}, compared to the ones in \cite{Kallosh:2010xz,GomezReino:2006dk} where only quartic corrections to the \K\, potential were studied, is that there are non-vanishing Christoffel symbols at $\BS=\BSbar=\BB=0$, due to the cubic corrections. In such a case, the corrections to the bosonic mass formula are not given by the curvature, but by the fourth derivative of the \K\, potential \cite{Kallosh:2010xz}
\be
\Delta{\cal M}_{i\jb}= -K_{i\jb k \bar l} \bar F^k F^{\bar l} \li.
\label{mass2}\ee
Moreover, we also find a holomorphic-holomorphic contribution to the mass matrix for the inflaton 
\be
\Delta{\cal M}_{ij}= -K_{i j k \bar l} \bar F^k F^{\bar l} \li.
\ee
We therefore find that with \rf{Delta}, there are non-vanishing Christoffel symbols as well as relevant fourth derivatives of the \K\, potential, defining the mass corrections due to the $c_i$ through
\be
K_{S\bar S S\bar S}= -4 c \, , \quad \Gamma_{\Phi\Phi}^S= -{1\over 2} \bar{c}_1\, , \quad \Gamma_{\bar{\Phi}\bar{\Phi}}^{\bar{S}}= -{1\over 2} c_1 \, , \quad K_{\Phi\bar \Phi S\bar S}=-K_{\Phi \Phi S\bar S} = -{1\over 2} c_2 .
\label{geom}\ee
It follows, according to (\ref{mass1}-\ref{geom}), that there is an additional contribution from the fourth derivatives of the \K\, potential, and from the Christoffel symbols (related to the third derivatives of the \K\, potential), to the masses of the bosons and the fermions. One also finds a contribution to the off-diagonal mass matrix ${\cal M}_{\Phi \Phi}= - {\cal M}_{\Phi \bar \Phi}$ for the inflaton, as explained in \cite{Kallosh:2010xz}. Therefore, the mass of the sinflaton has twice the value associated to $K_{\Phi\bar \Phi S\bar S}$.

With regard to the interpretation of the parameters $c$, $c_1$ and $c_2$, we have to keep in mind that the consistent limit where these parameters tend to infinity exists only when $F^\phi$ is a functional of fermions, as will be shown in Sec. \ref{sec.heavyInflatino}, and therefore has no vacuum expectation value. Also note that the interpretation of the parameters $c$, $c_1$ and $c_2$ going to infinity is associated to the masses of the particles that fall out of the spectrum. However, it can also be given in terms of the components of the curvature of the moduli space, going to infinity according to eq. \rf{curv}.

Instead of performing the analysis of the effect of the curved moduli space on the masses, we can use the action formula for the D-term action 
\be
\mathcal{L}_{K} = \int d^2 \theta\, d^2 \bar{\theta} \, {K(\BS, \BSbar; \BPhi , \BPhibar ) } \, , \qquad K= K_{\rm flat} + \Delta K,
\label{super}\ee
and compute all mass corrections due to $\Delta K$ directly.
Using \rf{super} we see that the corrections to the \K\, potential in \rf{Delta} lead to contributions to the masses of the sgoldstino $s$, inflatino $\chi^\phi$ and sinflaton $b$:
\be
- \mathcal{L}_{\rm mass}^{\rm sgoldst}= s\bar s \, c \, | 2 F^s|^2,
\label{mc}\ee 
\be
- \mathcal{L}_{\rm mass}^{\rm inflatino}= - \frac{1}{4} \Big (c_1F^s (\bar{\chi}^\phi)^2 + \bar c_1\bar F^s({\chi}^\phi)^2 \Big),
\label{mc1}\ee
\be
- \mathcal{L}_{\rm mass}^{\rm sinflaton}= b^2 c_2 | F^s|^2. 
\label{mc2}\ee 
Here we have used the fact that the bosonic part of $F^\phi$ is vanishing in the limit where $c$ and $c_1$ tend to infinity, see Sec. \ref{sec.heavyInflatino}. Because of this, we discard the term ${1\over 2} \chi^s \chi^\phi \bar F^\phi$ to the mass term in \rf{mc1}. The mass formulae in (\ref{mc}-\ref{mc2}) are consistent with the geometric analysis above.

The $c$-dependent terms are functions of component fields
\be
\mathcal{L}_{\Delta K} = \int d^2 \theta\,d^2 \bar{\theta} \, {\Delta K(\BS, \BSbar; \BPhi , \BPhibar ) } = c \, \mathcal{L}_{K}^c + c_1 \, \mathcal{L}_{K}^{c_1} + \overline{c_1 \, \mathcal{L}_{K}^{\bar c_1}} + c_2 \, \mathcal{L}_{K}^{c_2}.
\label{actions}\ee
The existence of the limits $c \rightarrow \infty$, $c_1\rightarrow \infty$ and $c_2\rightarrow \infty$ requires each of the equations
\be
\mathcal{L}_{K}^c =0\, ,\qquad \mathcal{L}_{K}^{c_1} =0\, , \qquad \mathcal{L}_{K}^{c_2} =0 \, ,
\label{EQ}\ee 
to be satisfied individually, with $c, c_2\in \mathbb{R}$ and $c_1\in \mathbb{C}$. We will begin by engaging only the term $- c {(\BS \BSbar)^2}$, which will permit us to send the mass of the sgoldstino to infinity through $c \rightarrow \infty$, at fixed values of $c_1, c_2$, and we will relate the resulting model to the one with the nilpotent superfield by requiring that
\be
\mathcal{L}_{K}^c =0\,. 
\label{EQ1}\ee 
This problem was solved in \cite{Kallosh:2015pho}. Here, we will propose an alternative way to derive and confirm that result, which will help us with the two other cases. 

Subsequently, we will add the correction $c_1 {\BS \BB^2} +\bar c_1 \BSbar \BB^2$, which gives rise to a mass term for the inflatino. This will permit us to send the masses of the sgoldstino and the inflatino to infinity with $c \rightarrow \infty$ and $c_1 \rightarrow \infty$, at fixed values of $ c_2$. We will require that
\be
\mathcal{L}_{K}^c =0\, ,\qquad \mathcal{L}_{K}^{c_1} =0\, ,
\label{EQ2}\ee 
and relate the resulting model to the one with the so-called `relaxed' constraint \cite{Komargodski:2009rz,Kahn:2015mla,Ferrara:2015tyn}, where the sinflaton field $b$ remains an independent field. 

Finally, we will also engage the term $c_2\BS\BSbar \BB^2$, leading to a mass term for the sinflaton, and relate the resulting model to the one with the orthogonal nilpotent constraints \cite{Komargodski:2009rz,Kahn:2015mla,Ferrara:2015tyn}, where the sinflaton field $b$ also becomes a dependent field. Altogether, we will find orthogonal nilpotent multiplet models in the limit of large masses --- for the sgoldstino and the inflatino, as well as the sinflaton. The limit of 
\be
c \rightarrow \infty\, , \qquad c_1\rightarrow \infty \, , \qquad c_2\rightarrow \infty
\label{limit}\ee
requires that all eqs. in \rf{EQ} are satisfied.

\subsection{The sgoldstino mass term}
In \cite{Kallosh:2015pho}, we presented an explicit expression for 
\be
\mathcal{L}_{K}^c = - c \int d^2 \theta\,d^2 \bar{\theta} \, {(\BS \BSbar)^2}
\label{action1}\ee
through computing the D-term of the superfield ${(\BS \BSbar)^2}$. By direct inspection of the explicit expression given in eq. (3.15) in \cite{Kallosh:2015pho} we concluded that it is {\it necessary and sufficient} to require that $s= {(\chi^s)^2\over 2 F^s}$ for that D-term to vanish, under condition that $F^s\neq 0$. We may now reformulate this requirement as follows. The superfield $(\BS \BSbar)^2$ is a product of a chiral and an anti-chiral superfield:
\be \label{eq.ZZ}
(\BS \BSbar)^2= Z\bar Z: \qquad {\bf Z}= \BS^2,\quad{\bf \bar Z}= \BSbar^2.
\ee
The only solution, invariant under global supersymmetry transformations, for which $\left.{\bf Z}\bar{\bf Z}\right|_D$ vanishes, is the constant chiral superfield ${\bf Z}=z_0$, which does not depend on $(x, \theta)$\footnote{For a proof of this statement, see appendix \ref{app.Dterms}.}. Let us apply this to the case of \rf{eq.ZZ}, with $\BS$ as given in \rf{S}. The $\theta$ and $\theta^2$ components of $\bf Z$ have to vanish, which implies
\be
2 s F^s - (\chi^s)^2 =0\, , \qquad s\chi^s=0.
\ee
Two types of solutions exist, depending on whether $F^s=0$ or $F^s\neq 0$. The first one is
\be
F^s=0\, , \qquad \chi^s=0\, , \qquad s=s \, : \qquad z_0= s^2,
\ee
whereas the second requires a strictly non-vanishing $F^s$ (but otherwise not restricted):
\be
F^s\neq 0\, , \qquad s={(\chi^s)^2\over 2 F^s} \, : \qquad z_0= s^2 =\Big ( {(\chi^s)^2\over 2 F^s} \Big )^2=0.
\label{our}\ee
Consistency of the second type of solutions requires that 
\be
F^s\neq 0 \, , \qquad z_0=0 \, \qquad \Rightarrow \qquad {\bf Z}={\BS}^2=0.
\ee
Thus we have given an explicit proof of the uniqueness of the requirement that ${\BS}^2=0$ for the D-term of ${(\BS \BSbar)^2}$ to vanish, under condition that $F^s\neq 0$. This is in agreement with the earlier derivation of the same result in \cite{Kallosh:2015pho}.

In conclusion, starting with linearly realized supersymmetry with an unconstrained, chiral superfield $\BS$, if $F^s\neq 0$ we find that the limit when the mass of the sgoldstino $m^2_s \sim c$ tends to infinity exists under the unique condition that the superfield $\BS$ is nilpotent.

\subsection{The inflatino mass term}
With the choice $c_1= a_1+i b_1$, we can present the inflatino mass term in the \K\, potential as
\be
a_1 {(\BS+ \BSbar) \BB^2} + b_1i\, (\BS- \BSbar) \BB^2.
\ee
That is, if we require the existence of the limit $a_1\rightarrow \infty$, the D-term $\left.{(\BS+ \BSbar) \BB^2}\right|_{D}$ must vanish, and similarly for the limit $b_1\rightarrow \infty$, the D-term $\left. {(\BS- \BSbar) \BB^2}\right|_D$ must vanish. These two conditions can be satisfied only if the D-term $\left. \BS {\bf B}^2\right|_D$ vanishes.

The goal is to find the necessary and sufficient conditions for the D-term of the superfield $\BS {\bf B}^2$ to vanish. The corresponding expression for the D-term has the following form: 
\begin{gather}\begin{aligned}
{\BS{\bf B}^2}|_D=& F^s\left[ib\bar{F}^\phi+\frac{1}{4}(\bar{\chi}^\phi)^2\right]+\\
&+\frac{1}{2}s\left[-b\square b+|F^\phi|^2+(\partial_\mu\varphi)(\partial^\mu\varphi)+\frac{i}{2}(\partial_{\mu}\chi^{\phi})\sigma^{\mu}\bar{\chi}^{\phi}-\frac{i}{2}\chi^{\phi}\sigma^{\mu}\partial_{\mu}\bar{\chi}^{\phi}\right]+\\
&+i(\partial_{\mu}s)\left[b(\partial_{\mu}\varphi)+\frac{1}{4}\chi^{\phi}\sigma^{\mu}\bar{\chi}^{\phi}\right]-\frac{1}{4}(\square s)b^2+\\
&+\frac{1}{2}\left[-(\chi^s\chi^{\phi})\bar{F}^{\phi}+\chi^s\sigma^{\mu}(\partial_{\mu}\bar{\chi}^{\phi})b+i\chi^s\sigma^{\mu}\bar{\chi}^{\phi}\partial_{\mu}\varphi \right]-\frac{1}{2}(\partial_{\mu}\chi^s)\sigma^{\mu}\bar{\chi}^{\phi}b.
\label{explicit}\end{aligned}\end{gather}
Expression \rf{explicit} includes the inflatino mass term $c_1F^s (\bar{\chi}^\phi)^2$, present for $\langle F^s\rangle\neq 0$, and in addition there exists a mixing term $-\frac{1}{2}c_1 (\chi^s\chi^{\phi})\bar{F}^{\phi}$. However, as we we will confirm below, the limit of $c, c_1$ to infinity requires that $\langle\bar F^\phi\rangle=0$. Therefore, we can discard the mixing term in this limit.

A direct inspection of expression \rf{explicit}, and a search for the conditions required to make it zero, is more complicated than in the case with a correction $- c {(\BS \BSbar)^2}$. A more effective approach is to look for a manifestly supersymmetric condition on some combination of the superfields involved, which would lead to the condition
\be
{\BS{\bf B}^2}|_D=0.
\ee
We require that
\be
\bar D_{\dot \alpha}({\BS{\bf B}})=0.
\label{chiral} \ee
It follows from \rf{chiral}, using the distributive property of the spinorial derivative, that
\be
\bar D_{\dot \alpha}({\BS{\bf B}^2})= {\BS} \bar D_{\dot \alpha}{{\bf B}^2}= 2 \Big ( {\BS} (\bar D_{\dot \alpha}{\bf B} )\Big ) \BB=0.
\label{proof}\ee
Here, we have shown that by requiring that $\bar D_{\dot \alpha}({\BS{\bf B}})=0$, one finds that the D-term for ${\BS{\bf B}^2}$ vanishes, since it is a chiral superfield.

It was shown in \cite{Komargodski:2009rz,Kahn:2015mla,Ferrara:2015tyn} that the condition \rf{chiral} requires that $\chi^\phi$ and $F^\phi$ are functions of the fields of the components of the nilpotent multiplet ${\BS}$, rather than independent fields. The explicit expressions are complicated. However, it follows from our argument in \rf{proof} that once the correct substitutions for $\chi^\phi$ and $F^\phi$, which solves \rf{chiral}, are inserted in \rf{explicit}, the right hand side of \rf{explicit} vanishes for arbitrary values of $\vp$ and $b$.

Concluding, the only known (nonconstant) superfields for which the D-term vanishes are either ${\bf Z}+\bar {\bf Z}$, with ${\bf Z}$ chiral, or linear ones, in the case of a real D-term. In our case, we have shown that $ {\BS}{\bf B}^2={\bf Z}$, and we know from \cite{Komargodski:2009rz,Kahn:2015mla,Ferrara:2015tyn} that there is a consistent solution. Since the linear multiplet case does not apply for the complex, general superfield $ {\BS}{\bf B}^2$, it means that the requirement \rf{chiral} is sufficient and necessary (see Sec. \ref{sec.heavyInflatino} for the local case).

\subsection{The sinflaton mass term}
Despite the fact that the D-term \rf{SSbarB2} is complicated, it reveals that there is a mass term for the sinflaton, $c_2 |F^s|^2b^2$, through
\begin{gather}\begin{aligned}\label{SSbarB2}
				\left. {\BS\BSbar}{\bf B}^2\right|_D =& \frac{1}{2}s\bar{s}\left[-b\square b+|F^\phi|^2+(\partial_\mu\varphi)(\partial^\mu\varphi)+\frac{i}{2}(\partial_{\mu}\chi^{\phi})\sigma^{\mu}\bar{\chi}^{\phi}-\frac{i}{2}\chi^{\phi}\sigma^{\mu}\partial_{\mu}\bar{\chi}^{\phi}\right]+\\
				&+\bigg[\frac{1}{4}s(\chi^{\phi})^2
				\bar{F}^s-\frac{1}{4}b^2 s\square \bar{s}-isb\bar{F}^s F^{\phi}-isb(\partial_{\mu}\vp)(\partial^{\mu}\bar{s})-\frac{i}{4}s\chi^{\phi}\sigma^{\mu}\bar{\chi}^{\phi}\partial_{\mu}\bar{s}-\\
				&\qquad-\frac{1}{2}s(\bar{\chi}^s\bar{\chi}^{\phi}) F^{\phi}+\frac{1}{2}s b (\partial_{\mu}\chi^{\phi})\sigma^{\mu}\bar{\chi}^{s}-\frac{i}{2}s\chi^{\phi}\sigma^{\mu}\bar{\chi}^s\partial_{\mu}\vp-\frac{1}{2}sb\chi^{\phi}\sigma^{\mu}\partial_{\mu}\bar{\chi}^s-\\
				&\qquad-\frac{i}{2}b^2 \chi^s\sigma^{\mu}\partial_{\mu}\bar{\chi}^s+\frac{1}{2}b\chi^s\sigma^{\mu}\bar{\chi}^{\phi}\partial_{\mu}\bar{s}+ib\bar{F}^s(\chi^s\chi^{\phi})+\text{h.c.}\bigg]+\\
				&+ \chi^s\sigma^{\mu}\bar{\chi}^{s}b\partial_{\mu}\vp+\frac{1}{2}(\chi^{\phi}\chi^s)(\bar{\chi}^{\phi}\bar{\chi}^s) +\frac{1}{2}b^2(\partial_{\mu}s)\partial^{\mu}\bar{s}+b^2|F^s|^2.
				\end{aligned}\end{gather}
From this explicit expression, one can find that the action vanishes only when the expression for $b$, in terms of the components of the nilpotent ${\BS}$ multiplet, is inserted. Instead of this complicated procedure, we will follow the strategy presented above. We note that the superfield of interest can be presented as a product of a chiral superfield $\BS {\bf B}$ (a constraint already implemented) times an anti-chiral one, $\BSbar\BB:
$
\be
\BS\BSbar{\bf B}^2= (\BS {\bf B}) (\BSbar{\bf B}) = {\bf Z} \bar {\bf Z}:\qquad {\bf Z} = \BS {\bf B}, \quad \bar {\bf Z} = \BSbar {\bf \bar B}.
\label{Dc2}\ee
The D-term of the product ${\bf Z} \bar {\bf Z}$ vanishes only if 
\be
{\bf Z} = { \BS {\bf B}}= z_0,
\ee
i.e., we have to determine whether $z_0$ vanishes, or not. With $ \BS {\bf B}$ chiral, we can apply the procedure in appendix \ref{app.Dterms} with
\be
{\bf Z} = {\BS {\bf B}} \\
= z^{\BS\BB}+\sqrt{2}\theta \chi^{\BS\BB}+\theta^2F^{\BS\BB}
\ee
for some scalar $z^{\BS\BB}$, fermion $\chi^{\BS\BB}$ and auxiliary field $F^{\BS\BB}$. From the expansion of $\BS\BB$ we have
\begin{align}\label{sbscalars}
z^{\BS\BB} = sb &= z_0,\\
\chi^{\BS\BB} = -\frac{i}{2} s \chi^{\phi}+b\chi^s &= 0, \\
F^{\BS\BB} = bF^s-\frac{i}{2}sF^{\phi}+\frac{i}{2}(\chi^s\chi^{\phi}) 	&= 0.
\label{aux1}\end{align}
From \rf{aux1} it follows that
\be
b={i\over {2F^s}} \Big (sF^{\phi}-(\chi^s\chi^{\phi}) \Big),
\ee
since $F^s\neq 0$. Consequently, we find
\be
sb= {{is}\over {2F^s}} \Big (sF^{\phi}-(\chi^s\chi^{\phi}) \Big)=0.
\ee
We have thus shown that the condition that the D-term of \rf{Dc2} vanishes, requires that the corresponding chiral superfield vanishes:
\be
{\bf Z} = {\BS {\bf B}}=0.
\ee
As we know, this condition, in addition to $\BS^2=0$, leads to constraints on the components of the inflaton multiplet, which fix the sinflaton $b$ to become a function of of the components of the nilpotent multiplet ${\BS}$.

This finalizes our proof that in the limit \rf{limit} when the masses of the sgoldstino, the inflatino and the sinflaton tend to infinity
\be
m^2_{s} \rightarrow \infty\, , \qquad m_{\chi^\phi} \rightarrow \infty\, , \qquad m^2_b \rightarrow \infty,
\ee
the linear model becomes the one with non-linearly realized supersymmetry and orthogonal nilpotent superfields.

\section{Local supersymmetry}\label{sec:local}
The local case with $- c{(\BS \BSbar)^2}$ was studied in \cite{Kallosh:2015pho}. We will here assume that $\BS$ is nilpotent and proceed with the new mass terms for the inflatino and the sinflaton, respectively. The multiplet calculus which we use below, in the form developed in \cite{FKVW}, is valid for local supersymmetry; all derivatives are supercovariant.

\subsection{Basic multiplet calculus }
To find a generalization to local supersymmetry of the expression for our actions in \rf{actions}, we apply multiplication laws for complex multiplets. A generic complex multiplet $\BC$ has the components
\begin{equation}
 \left\{{\cal C},{\cal Z},{\cal H},{\cal K},{\cal B}_\mu ,\Lambda ,{\cal D}\right\},
 \label{componentsComplexm}
\end{equation}
where ${\cal C}$, ${\cal H}$, ${\cal K}$ and ${\cal D}$ are complex scalars, and ${\cal Z}$ and $\Lambda $ are Dirac fermions. Moreover, a generic complex multiplet $\BC^3$, the product of two multiplets $\BC^1$ and $\BC^2$, has the following components\footnote{Compare with the general case in \cite{FKVW}, with $f({\cal C}^i)$: $f_1 = {\cal C}^2$, $f_2 = {\cal C}^1$ and $f_{12} = 1$.} \cite{FKVW}

\begin{gather}\begin{aligned}
{\cal C}^3 & = {\cal C}^1 {\cal C}^2,\\
{\cal Z}^3 & ={\cal C}^1{\cal Z}^2 + {\cal C}^2{\cal Z}^1,\\
{\cal H}^3 &= {\cal C}^1\H^2+{\cal C}^2\H^1-\frac{1}{2}\bar{\Z}^1P_L\Z^2-\frac{1}{2}\bar{\Z}^2P_L\Z^1 ,\\
{\cal K}^3 &= {\cal C}^1\mK^2+{\cal C}^2\mK^1-\frac{1}{2}\bar{\Z}^1P_R\Z^2-\frac{1}{2}\bar{Z}^2P_R\Z^1 ,\\
{\cal B}_\mu^3 &= {\cal C}^1\B^2+{\cal C}^2\B^1+\frac{i}{2}\bar{\Z}^1P_L\gamma_{\mu}\Z^2+\frac{i}{2}\bar{\Z}^2P_L\gamma_{\mu}\Z^1 ,\\
{\Lambda }^3&={\cal C}^1 \Lambda^2+{\cal C}^2 \Lambda^1 + \frac{1}{2} \left[\rmi\gamma _*\slashed{\cal B}^1+P_L{\cal K}^1+P_R{\cal H}^1-\slashed{\cal D}{\cal C}^1\right]{\cal Z}^2+\\
& \quad+\frac{1}{2} \left[\rmi\gamma _*\slashed{\cal B}^2+P_L{\cal K}^2+P_R{\cal H}^2-\slashed{\cal D}{\cal C}^2\right]{\cal Z}^1 ,\\
{\cal D}^3&={\cal C}^1{\cal D}^2+{\cal C}^2{\cal D}^1+
{{\cal K}^1{\cal H}^2+ {\cal K}^2{\cal H}^1\over 2} -{\cal B}^1\cdot {\cal B}^2- {\cal D}{\cal C}^1\cdot {\cal D}{\cal C}^2-\bar \Lambda ^1{\cal Z}^2-\bar \Lambda ^2{\cal Z}^1\\
&\quad -{\bar {\cal Z}^1\slashed{\cal D}{\cal Z}^2+ \bar {\cal Z}^2\slashed{\cal D}{\cal Z}^1\over 2} .
\label{prod}
\end{aligned}\end{gather}
Here, the supercovariant derivatives for the case of local supersymmetry are defined in \cite{FKVW}, for the global case they are simple derivatives.

\subsection{Real and (anti-) chiral fields}
The complex multiplet reduces to a \emph{real} multiplet when $C={\cal C}$ is real. Then $\cal Z$ and $\Lambda $ are Majorana, i.e. $(P_R{\cal Z})^C=P_L{\cal Z}$. Furthermore, ${\cal K}={\cal H}^*$ while ${\cal B}_\mu $ and ${\cal D}$ are real, yielding the components
\begin{equation}
 \left\{C,\,\zeta ,\,{\cal H},\, {\cal H}^*,\, B_\mu ,\,\lambda ,\, D\right\}.
 \label{realmultiplet}
\end{equation}
On the other hand, it reduces to a \emph{chiral} multiplet for $P_R{\cal Z}=0$, with the components \cite{FKVW}
\be \label{chiralrestrcomplex}
\left\{{\cal C},\,P_L{\cal Z} ,\,{\cal H},\, 0,\, \rmi{\cal D}_\mu {\cal C} ,0 ,0\right\}.
\ee

In our setting, with ${\BS}$ a chiral multiplet containing $\{s,P_L \Omega^s,F^s\}$ and $\BB$ the real multiplet given in \rf{eq.B}, with $\bf \Phi$ chiral and containing \mbox{$\{\vp+ib,P_L \Omega^\phi,F^\phi\}$}, the relevant expressions in terms of complex multiplets are \cite{FKVW}
\be
\begin{array}{ccccccccccc}
{\bf C} & = & \{ & \C, & \Z, & \H, & {\cal K}, & \B, & \L, & \D & \}\,,\\
{\bf S} &=& \{ &s,&-\rmi \sqrt{2}P_L\Omega^s,&-2F^s,&0,& \rmi \D_\mu s,&0,&0&\}\,,\\
{\bf \Phi} &=& \{& \vp+\rmi b,&-\rmi \sqrt{2}P_L\Op,&-2F^\phi,&0,& \rmi \D_\mu (\vp+\rmi b),&0,&0&\}\,,\\
{\bf \bar{\Phi}} &=& \{ &\vp-\rmi b,&\rmi \sqrt{2}P_R\Op,&0,&-2\bar{F}^{\phi},&- \rmi \D_\mu (\vp-\rmi b),&0,&0&\}\,,\\
{\bf B} &=& \{& b,&-\frac{1}{\sqrt{2}} \Op,& \rmi F^\phi,& -\rmi \bar{F}^{\phi},&\D_\mu \vp,&0,&0&\}\,.
\label{table}\end{array}
\ee
Here, $\vp$ and $b$ are real, while the other quantities are complex, and it is possible to check that the nilpotency condition \rf{eq.nilpS} requires that
\be \label{sgold}
s= {\bar \Omega^s P_L \Omega^s\over 2 F^s},
\ee
using the multiplet calculus \rf{prod}.

\subsection{The heavy inflatino model}\label{sec.heavyInflatino}
With the addition of the term $(c_1 {\BS}+\bar c_1 \BSbar){\bf B}^2$ to the \K\, potential, the D-term component of the superfield $c_1 {\BS}\BB^2 + \bar c_1 {\BSbar}\BB^2$ has the form $c_1 \mathcal{L}_{K}^1 + \overline{c_1 \mathcal{L}_{K}^1} $. In relation to this, the first question that can be asked is under which conditions the D-component of the superfield vanishes, without the complete superfield ${\BS}\BB^2$ vanishing. A tentative answer is that the only supersymmetric condition which may lead to such a result is based on the properties of the superfield $\BB$ being the sum of a chiral and an antichiral superfield, and we will see that if we assume ${\BS}\BB$ to be chiral, it is possible to deduce that all terms in the D-term of $c_1 {\BS}\BB^2 + \bar c_1 {\BSbar}\BB^2$ vanish. Our first step is therefore to impose the restrictions of \rf{chiralrestrcomplex} on the components of ${\BS}\BB$. They are set by \rf{prod} to be \cite{FKVW}
\begin{gather}\begin{aligned}\label{eq:SB}
\C^{\BS\BB} &= s b,\\
\Z^{\BS\BB} &= -\rmi b \sqrt{2}P_L\Omega^s - \frac{1}{\sqrt{2}} s \Op , \\
\H^{\BS\BB} &=-2F^s b +\rmi s F^\phi-\rmi\bar{\O}^s P_L \Op, \\
{\cal K}^{\BS\BB} &= -\rmi s \bar{F}^\phi, \\
\B^{\BS\BB} &= \rmi b \D_\mu s+s \D_\mu \vp- \frac12 \bar{\O}^s P_L \gamma_\mu \Op, \\
\L^{\BS\BB} &=\frac{1}{\sqrt{2}} \lp F^s+(\slashed{\D} s) \rp P_R \Op -\frac{1}{\sqrt{2}} \lp[\slashed{\D} (\vp-\rmi b)] + \bar{F}^\phi \rp P_L \O^s, \\
\D^{\BS\BB} &=i F^s \bar{F}^\phi - \rmi [\D^\mu (\vp-\rmi b)](\D_\mu s)-\frac{\rmi}{2} \bar{\Omega}^s P_L \slashed{\D} \Op - \frac{\rmi}{2} \bar{\Omega}^\phi \slashed{\D} P_L \O^s,
\end{aligned}\end{gather}
where the condition on $\Lambda^{\BS\BB}$ splits into two since the left and right projections must vanish independently. Of these, the right projections gives
\be \label{inflatino} 
 P_R \Op = \left[\slashed{\D} (\vp-\rmi b)\right]P_L {\O^s\over F^s}.
\ee
The condition on ${\cal D}^{\BS \BB}$ then gives
\be \label{aux}
 \bar{F}^\phi = \left[\D^\mu (\vp-\rmi b)\right]{\D_\mu s\over F^s}+\frac{1}{2 F^s} \bar{\Omega}^s \slashed{\D} P_R\Op + \frac{1}{2 F^s} \bar{\Omega}^\phi P_R\slashed{\D} \O^s,
 \ee
the solution of which can be obtained by substituting in the expressions given in \rf{sgold} and \rf{inflatino}, as derived in appendix \ref{app.derivations}:
 \be
\bar{F}^\phi = \left(\D_\nu{\bar \Omega^s\over F^s} \right) \g^\mu \g^\nu P_L \left({\Omega^s\over F^s} \right) \D_\mu (\vp -i b) + {s\over F^s} \D^2 (\vp -i b).
\label{Fphi}\ee
In total, this satisfies all of the conditions for $\BS\BB$ to be a chiral multiplet; the $P_R {\cal Z}^{\BS\BB}$ vanishes etc. We have derived the expressions for the inflatino and for the auxiliary field in the inflaton multiplet as functions of $\vp$, $b$, $\Omega^s$ and $F^s$. From \rf{Fphi} is is clear that $F^\phi$ is a functional of fermions and does not contribute to the bosonic sector of the theory. The expression in 
\rf{Fphi} is a supergravity generalization of the corresponding expression in \cite{Komargodski:2009rz,Kahn:2015mla}. 

\subsubsection*{Proof of the chirality of the ${\BS}{\bf B}^2$ superfield} 
We would like to compute a complete expression for ${\cal L}^1_K$, i.e. we need to derive the supergravity D-term action of the $\BS \BB^2$ superfield, a supergravity version of \rf{explicit}. But since we know that such a D-term is vanishing in the case of global supersymmetry, we may want to prove the analogous result in the local case instead. The simplest way to do this is to use the multiplet calculus presented above with the concept of a covariant superspace derivative, $\bar {\cal D}_{\dot \alpha}$ \cite{FKVW}, so that $\bar {\cal D}^3=0$. Such a derivative, like in the global case, has a distributive property, so since we have 
\be
\bar {\cal D}_{\dot \alpha}({\BS{\bf B}})=0,
\label{chiral1} \ee
it follows that
\be
\bar {\cal D}_{\dot \alpha}({\BS{\bf B}^2})= {\BS} \bar {\cal D}_{\dot \alpha}{{\bf B}^2}= 2 \Big ( {\BS} (\bar {\cal D}_{\dot \alpha}{\bf B} )\Big ) \BB=0.
\label{proof1}\ee
An equivalent way to prove that $\BS \BB^2$ is a chiral superfield is to show that 
\be
P_R \Z^{\BS\BB^2} =0. \label{right}
\ee
We use the second line in \rf{eq:SB}, act on it with the projector $P_R$:
\be
P_R {\cal Z}^{\BS\BB^2} ={\cal C}^{\BS\BB} P_R{\cal Z}^\BB + {\cal C}^\BB P_R{\cal Z}^{\BS\BB},
\ee
and note that these two terms can be given on the form $P_R {\cal Z}^{\BS\BB^2} =2\, {\cal C}^\BB P_R{\cal Z}^{\BS\BB}$. Hence, for the unconstrained $b$, the vanishing of this term is possible only if ${\BS{\bf B}}$ is chiral.

\subsection{The heavy sinflaton model}
Adding the term $-c_2\BS\BSbar \BB^2 $ to the \K\ potential, after the addition of the inflatino mass term, and requiring the existence of the limit $c_2\rightarrow \infty$, we find
\be
{\BS{\bf B}}=0,
\ee
as in the global case.

The easy way to find the constraint solution for $b$ is to solve the condition $\H^{\BS\BB}=0$, with $\H^{\BS\BB}$ as given in \rf{eq:SB}. This defines $b$ as
\be
b= {1\over 2F^s} \Big( \rmi s F^\phi-\rmi\bar{\O}^s P_L \Op\Big)={1\over 2\bar F^s} \Big( - \rmi \bar s \bar{F}^\phi+\rmi\bar{\O}^s P_R\Op\Big),
\label{b}\ee
where the latter equality is given by the reality of $b$, giving the solution in a form more suitable for substituting in the expressions in \rf{inflatino} and \rf{aux}. Note that $sb=0$. Finally, we obtain
\begin{gather}\begin{aligned}
b&= \frac{i}{4}\Bigg[\frac{\bar{\Omega}^s}{\bar{F}^s}\g^\mu P_L\frac{\Omega^s}{F^s}-\frac{\bar{s}}{\bar{F}^s}\left({\cal D}_\nu\frac{\bar{\Omega}^s}{F^s}\right)\g^\mu\g^\nu P_L \frac{\Omega^s}{F^s}-\\
&\qquad\quad-\frac{s\bar{s}}{2|F^s|^2}\left({\cal D}_{\nu}\frac{\bar{\Omega}^s}{\bar{F}^s}\right)(\g^{\mu\nu\rho}+\g^\mu\eta^{\nu\rho}) P_L\left({\cal D}_{\rho}\frac{\Omega^s}{F^s}\right) -c.c. \Bigg]{\cal D}_\mu\vp,
\end{aligned}\end{gather}
as derived in appendix \ref{app.derivations}, where a comparison with previous work also is to be found. This expression was presented in the case of global supersymmetry, in a different notation, in \cite{Komargodski:2009rz,Kahn:2015mla}.

\section{Summary}
In this paper, we have shown that by sending the masses of the sgoldstino, the inflatino and the sinflaton to infinity, it is possible to derive the action with orthogonal nilpotent supermultiplets in supergravity. The mechanism is analogous to the one we have developed for the case of one nilpotent multiplet in \cite{Kallosh:2015pho}. Our new results are in agreement with the description of orthogonal nilpotent models in \cite{Ferrara:2015tyn} and are based on a proposal there as to how to modify the \K\, potential for this purpose. We have used the multiplet tensor calculus for general supermultiplets developed in \cite{FKVW} to consistently define a general class of models with constrained superfields.

\section*{Acknowledgements}
We are grateful to Sergio Ferrara, Jesse Thaler, Antoine Van Proeyen and Timm Wrase for stimulating discussions and collaborations on related projects. This work is supported by the SITP and by the NSF Grant PHY-1316699. BM is supported by the Netherlands Organisation for Scientific Research (NWO). AK is supported by the K.~A.~Wallenberg Foundation.

\appendix

\section{Comparing with earlier work} \label{app.EarlierWork}
In \cite{Dudas:2011kt} the following choice was made for the total action
\begin{gather}\begin{aligned}
\mathcal{L} &= \int d^4 \theta \Big [ {\BS^\dagger \BS + \BPhi^\dagger \BPhi} - c {(\BS^\dagger \BS)^2 } + {M\over 4f^2} \, { \BS^\dagger \BS (D^\alpha \BPhi D_\alpha \BPhi + \bar D^{\dot \alpha} \BPhi^\dagger \bar D_{\dot \alpha} \BPhi^\dagger)}\Big] +\\
&\quad+ \int d^2 \theta f \, {\BS} +\rm{h.c.}
\label{dudas}
\end{aligned}\end{gather}
This action was called a `microscopic theory' underlying the constrained superfields \rf{relaxed}. The component Lagrangian at zero-momentum for heavy fields $s$ and $\chi^\phi$ was computed, as in a simpler case of only a term ${(\BS^\dagger \BS)^2 }$ in \cite{Komargodski:2009rz}, and the field equations for $\chi^\phi$ and $F^\phi$ were derived. For $\chi^\phi$, the equation which follows from this `microscopic theory' is in agreement with the one following from the solutions of the superfield constraint \rf{relaxed}. However, the one for the auxiliary field is of the form
\be
F^\phi= -{2M\over f^2} s \bar s \Box (\vp +i b).
\label{Fdud} \ee
This is totally different from the expression expected from the superfield constraint solution where, according to \cite{Komargodski:2009rz},
 \be
 F^\phi= -\partial_\nu \Big({\bar \chi^s\over \bar F^s} \Big) \bar \sigma^\mu \sigma^\nu \Big({\bar \chi^s\over \bar F^s} \Big) \partial_\mu (\vp +i b) + \frac{1}{2}\Big({\bar \chi^s\over \bar F^s} \Big)^2 \partial^2 (\vp +i b).
\label{auxKS} 
\ee
The first term in \rf{auxKS} is missing in \rf{Fdud}, and the second term in \rf{auxKS} significantly differs from \rf{Fdud}. The authors qualified this discrepancy as a `puzzle' which they have never resolved. 

Now we see that by making a choice\footnote{It is possible that with the choice \rf{dudas}, one may find an agreement with the superfield constraints by taking the limit to $c\rightarrow \infty, M\rightarrow \infty$, but this requires a separate investigation.} of the linear supersymmetry model as in eqs. (\ref{flat}, \ref{Delta}), with a non-vanishing $F^s$, we have proven the consistency of both the `relaxed' \rf{relaxed} as well as the orthogonal \rf{orth} nilpotent constraints. We did not use the IR approximation for heavy fields, as in \cite{Komargodski:2009rz,Dudas:2011kt}, but instead studied the existence of the formal limit when $c, c_1$ and $c_2$ tend to infinity, and found a complete agreement with the superfield constraints.

\section{A theorem on vanishing D-terms}\label{app.Dterms}
For a chiral superfield $\bf Z$ the D-term is 
\be\label{intDterm}
\int d^4x \int d^2\theta d^2\bar{\theta} \ \ {\bf Z}\bar{\bf Z}.
\ee
Here, we will show that the vanishing of this term implies that a solution ${\bf Z}(x, \theta)$ that is invariant under global supersymmetry transformations must be an $(x, \theta)$-independent constant $z_0$:
\be\label{thm}
\int d^4x \int d^2\theta d^2\bar{\theta} \ \ {\bf Z}\bar{\bf Z} = 0 \qquad \Rightarrow \qquad {\bf Z}=z_0.
\ee

A chiral field can be expanded in 2-component spinors as
\be
{\bf Z}(\theta, y) = z(y)+\sqrt{2}\theta \chi^z(y)+\theta^2F^z(y),
\ee
with $y^{\mu} = x^{\mu}+i\theta\sigma^{\mu}\bar{\theta}$, and the D-term of ${\bf Z}\bar{\bf Z}$ given by
\be \label{Dtermzz}
\left.{\bf Z}\bar{\bf Z}\right|_D = F^z\bar{F}^z+(\partial_{\mu}z)(\partial^{\mu}\bar{z})-i\chi^z\sigma^{\mu}\partial_{\mu}\bar{\chi}^z,
\ee
up to total derivatives.

The transformations of ${\bf Z}$ under global supersymmetry are given by
\ba
\delta_{\epsilon}z &=& \sqrt{2}\epsilon \chi^z,\label{scalartransform}\\
\delta_{\epsilon}\chi^z &=& i\sqrt{2}\sigma^{\mu}\bar{\epsilon}\partial_{\mu}z+\sqrt{2}\epsilon F^z,\label{fermiontransform}\\
\delta_{\epsilon}F^z &=& i\sqrt{2}\bar{\epsilon}\bar{\sigma}^{\mu}\partial_{\mu}\chi^z.\label{auxtransform}
\ea
We require that the D-term (\ref{intDterm}) vanishes for a solution ${\bf Z}$ that is invariant under global supersymmetry transformations. The variation of the scalar and auxiliary components (\ref{scalartransform},\ref{auxtransform}) vanishes if and only if the fermionic component $\chi^z$ is zero everywhere. In particular, we require the variation of $\chi^z$ (\ref{fermiontransform}) to vanish, which gives
\be\label{fermionvariationvanish}
i\epsilon\sigma^{\mu}\bar{\epsilon}\partial_{\mu}z =-\epsilon^2 F^z\quad\Rightarrow \quad(\partial_{\mu}z)(\partial^{\mu}\bar{z}) = 2F^z\bar{F}^z.
\ee

From the requirement that \rf{Dtermzz} vanishes, it now follows that
\be
{\bf Z} = (z,\chi^z,F^z) \\
= (z_0,0,0), \qquad \partial_{\mu}z_0 =0,
\ee
or in other words, $\bf Z$ is a superfield where the only nonzero part is a constant scalar. 
This concludes the proof of (\ref{thm}).

\section{Derivations of $\bar{F}^\phi$ and $b$}\label{app.derivations}
For the derivation of the expressions for $\bar{F}^\phi$ and $b$, we use relations for Majorana spinors in four dimensions. For example, \cite{Freedman:2012zz}
\be \label{eq.initRel}
\bar{\Omega}\Omega'=\bar{\Omega}'\Omega,\quad \bar{\Omega}\g^\mu\Omega'=-\bar{\Omega}'\g^\mu\Omega
\ee
gives 
\begin{gather}\begin{aligned}\label{eq.swPhi}
&P_R\Omega^\phi=\left[\slashed{\cal D}(\vp-ib)\right]P_L\frac{\Omega^s}{F^s}\quad\Rightarrow\\
&\bar{\Omega}^\phi P_R=-\frac{\bar{\Omega}^s}{F^s}P_L{\slashed{\cal D}}(\vp-ib).
\end{aligned}\end{gather}
Other useful relations include \cite{Freedman:2012zz}
\begin{subequations}\label{eq.rels}
\begin{align}
&\bar{\Omega}\g^\mu\g^\nu\Omega=\bar{\Omega}\Omega\eta^{\mu\nu} \label{eq.id},\\
&\bar{\Omega}'\g^\mu\g^\nu\Omega= \bar{\Omega}\g^\nu\g^\mu\Omega'
\end{align}
\end{subequations}
in combination with \rf{sgold}, as well as \cite{Freedman:2012zz}
\begin{gather}\begin{aligned}\label{eq.Fi3}
(\bar{\Omega}P_L\Omega')(\bar{\Omega}P_L\Omega'')&=-\frac{1}{2}(\bar{\Omega}P_L\Omega)(\bar{\Omega}'P_L\Omega''),\\
\g^\mu\g^\nu\g^\rho&=\g^{\mu\nu\rho}+2\g^{[\mu}\eta^{\nu]\rho}+\g^\rho\eta^{\mu\nu}.
\end{aligned}\end{gather}

\subsubsection*{The expression for $\bar{F}^\phi$}
For $\bar{F}^\phi$, we have \rf{aux}:
\begin{gather}\begin{aligned}
&\bar{F}^\phi=\left[{\cal D}_\mu(\vp-ib)\right]\frac{{\cal D}^\mu s}{F^s}+\frac{1}{2F^s}\bar{\Omega}^\phi P_R{\slashed{\cal D}}\Omega^s+\frac{1}{2F^s}\bar{\Omega}^s{\slashed{\cal D}}P_R\Omega^\phi=\\
&\stackrel{(\ref{eq.swPhi},\li\ref{eq.id})}{=}\frac{1}{2F^s}\left[{\cal D}_\mu(\vp-ib)\right]\left[{\cal D}_\nu \frac{\bar{\Omega}^s}{F^s}P_L\g^\mu\g^\nu\Omega^s-\frac{\bar{\Omega}^s}{F^s}P_L\g^\mu\slashed{\cal D}\Omega^s
\right]+\frac{1}{2F^s}\bar{\Omega}^s{\slashed{\cal D}}P_R\Omega^\phi=\\
&=\frac{1}{2F^s}\left[{\cal D}_\mu(\vp-ib)\right]\left({\cal D}_\nu \frac{\bar{\Omega}^s}{F^s}\right)P_L\g^\mu\g^\nu\Omega^s+\frac{1}{2F^s}\bar{\Omega}^s{\slashed{\cal D}}P_R\Omega^\phi=\\
&\stackrel{(\ref{eq.swPhi},\li\ref{eq.rels})}{=}\frac{1}{F^s}\left({\cal D}_\nu \frac{\bar{\Omega}^s}{F^s}\right)P_L\g^\mu\g^\nu\Omega^s{\cal D}_\mu(\vp-ib)+\frac{s}{F^s}{\cal D}^2(\vp-ib),\nonumber
\end{aligned}\end{gather}
giving the final result
\be \label{eq.Ffinal}
\bar{F}^\phi=\left({\cal D}_\nu \frac{\bar{\Omega}^s}{F^s}\right)P_L\g^\mu\g^\nu\frac{\Omega^s}{F^s}{\cal D}_\mu(\vp-ib)+\frac{s}{F^s}{\cal D}^2(\vp-ib).
\ee

\subsubsection*{The expression for $b$}
For $b$, we have \rf{b}:
\begin{gather}
\begin{aligned}
b&=\frac{i}{2\bar{F}^s}\left(-\bar{s}\bar{F}^\phi+\bar{\Omega}^sP_R\Omega^\phi\right)=\\\nonumber
&\stackrel{(\ref{eq.swPhi},\ref{eq.Ffinal})}{=}\frac{i}{2|F^s|^2}\left[\stackrel{{\blue (a)}}{-s\bar{s}{\cal D}^2}-\bar{s}\left({\cal D}_\nu\frac{\bar{\Omega}^s}{F^s}\right)\stackrel{{\blue (b)}}{\g^\mu\g^\nu} P_L \Omega^s {\cal D}_\mu+\stackrel{{\blue (c)}}{\bar{\Omega}^s\g^\mu P_L\Omega^s{\cal D}_\mu}\right](\vp-ib),
\end{aligned}
\end{gather}
which can be solved through iterations, with the extra terms introducing more derivatives on $\Omega^s$. Since each type of spinor only has two degrees of freedom ($s P_L\Omega^s=0$), the first (a) and second (b) terms only get corrected through the third (c) being fed back into $b$. Importantly, the derivatives have to act on the spinors in the $b$ in a way that does not render the expression to zero. For (a) and (b), we have
\begin{subequations}
\begin{gather}\begin{aligned}\label{eq.1inb}
&\frac{i}{2|F^s|^2}[-s\bar{s}{\cal D}^2](\vp-ib)\rightarrow\\
&-\frac{is\bar{s}}{2|F^s|^2}\left[{\cal D}^2\vp+\frac{1}{|F^s|^2}({\cal D}^\nu\bar{\Omega}^s)\g^\mu P_L({\cal D}_\nu\Omega^s){\cal D}_\mu\vp\right]
\end{aligned}\end{gather}
\begin{gather}\begin{aligned}\label{eq.2inb}
&\frac{i}{2|F^s|^2}\bigg[-\bar{s}\left({\cal D}_\nu\frac{\bar{\Omega}^s}{F^s}\right)\g^\mu\g^\nu P_L \Omega^s {\cal D}_\mu\bigg](\vp-ib)\rightarrow\\
&-\frac{i\bar{s}}{2|F^s|^2}\left({\cal D}_\nu\frac{\bar{\Omega}^s}{F^s}\right)\g^\mu\g^\nu P_L \Omega^s \bigg[{\cal D}_\mu\vp+\frac{1}{2|F^s|^2}({\cal D}_\mu\bar{\Omega}^s)\g^\rho P_L\Omega^s{\cal D}_\rho\vp\bigg],
\end{aligned}\end{gather}
whereas the third term (c) receives corrections from both the second (b) and the third (c) terms:
\begin{gather}\begin{aligned}\label{eq.3inb}
&\frac{i}{2|F^s|^2}\bar{\Omega}^s\g^\mu P_L\Omega^s{\cal D}_\mu(\vp-ib)\rightarrow\\
&\frac{i}{2|F^s|^2}\bar{\Omega}^s\g^\mu P_L\Omega^s\bigg[{\cal D}_\mu\vp-\frac{1}{2|F^s|^4}\left({\bar{\Omega}^s P_R{\cal D}_\mu\Omega^s}\right)\left({\cal D}_\nu\bar{\Omega}^s\right)\g^\rho\g^\nu P_L \Omega^s {\cal D}_\rho\vp+\\
&\qquad\quad\qquad\quad+\frac{1}{2|F^s|^2}\bar{\Omega}^s\g^\nu P_L\Omega^s{\cal D}_\mu{\cal D}_\nu(\vp-ib)+\frac{1}{2}\left({\cal D}_\mu\frac{\bar{\Omega}^s}{\bar{F}^s}\g^\nu P_L\frac{\Omega^s}{F^s}\right){\cal D}_\nu(\vp-ib)\bigg].
\end{aligned}\end{gather}
\end{subequations}
Here, the terms in the last line again get corrected by $b$, but only by the (uncorrected) term (c) due to the amount of spinors in the other terms. Through simplifications using (\ref{eq.rels}-\ref{eq.Fi3}) and flips as in \eqref{eq.swPhi}, it it possible to see that the third term in \eqref{eq.3inb} cancels the contributions from \eqref{eq.1inb}. In a similar way, the second terms in \eqref{eq.2inb} and \eqref{eq.3inb} become
\be \label{eq.middle2}
\frac{is\bar{s}}{4|F^s|^4}({\cal D}_\nu \bar{\Omega}^s) (\g^\mu\g^\rho\g^\nu-\g^\nu\g^\rho\g^\mu) P_L({\cal D}_\rho\Omega^s){\cal D}_\mu \vp\stackrel{\ref{eq.Fi3}}{=}-\frac{is\bar{s}}{2|F^s|^4}({\cal D}_\nu \bar{\Omega}^s) \g^{\mu\nu\rho} P_L({\cal D}_\rho\Omega^s){\cal D}_\mu \vp\li,
\ee
where $\Omega\g^{\mu\nu\rho}\Omega'=\Omega'\g^{\mu\nu\rho}\Omega$ \cite{Freedman:2012zz}. Meanwhile, the last term in \eqref{eq.3inb} is
\be\label{eq.3inb2}
\frac{i}{4|F^s|^2}\Bigg[s\left({\cal D}_\mu\frac{\bar{\Omega}^s}{\bar{F}^s}\right)\g^\nu\g^\mu P_R\Omega^s{\cal D}_\nu\vp + c.c.\Bigg]
\ee
and, through replacing the $b$ with the first term in \eqref{eq.3inb}
\begin{gather}\begin{aligned}\label{eq.lastTerm}
&-\frac{is\bar{s}}{8|F^s|^4}({\cal D}_\nu\bar{\Omega}^s)(\g^\rho\g^\nu\g^\mu+\g^\mu\g^\rho\g^\nu) P_L({\cal D}_\rho\Omega^s){\cal D}_\mu\vp\stackrel{\ref{eq.Fi3}}{=}\\
&=\frac{is\bar{s}}{4|F^s|^4}({\cal D}_\nu\bar{\Omega}^s)(\g^{\mu\nu\rho}-\g^\mu\eta^{\nu\rho}) P_L({\cal D}_\rho\Omega^s){\cal D}_\mu\vp\li.
\end{aligned}\end{gather}

In total we get the first term in \eqref{eq.3inb}, \eqref{eq.3inb2} with the first term in \eqref{eq.2inb}, and \eqref{eq.middle2} with \eqref{eq.lastTerm}:
\begin{gather}\begin{aligned}
b&=\frac{i}{2|F^s|^2}\bar{\Omega}^s\g^\mu P_L\Omega^s{\cal D}_\mu\vp-\left[\frac{i\bar{s}}{4|F^s|^2}\left({\cal D}_\nu\frac{\bar{\Omega}^s}{F^s}\right)\g^\mu\g^\nu P_L \Omega^s {\cal D}_\mu\vp+c.c.\right]-\\
&\quad-\frac{is\bar{s}}{4|F^s|^4}({\cal D}_\nu\bar{\Omega}^s)(\g^{\mu\nu\rho}+\g^\mu\eta^{\nu\rho}) P_L({\cal D}_\rho\Omega^s){\cal D}_\mu\vp=\\
&=\frac{i}{4}\Bigg[\frac{\bar{\Omega}^s}{\bar{F}^s}\g^\mu P_L\frac{\Omega^s}{F^s}-\frac{\bar{s}}{\bar{F}^s}\left({\cal D}_\nu\frac{\bar{\Omega}^s}{F^s}\right)\g^\mu\g^\nu P_L \frac{\Omega^s}{F^s}-\\
&\qquad\quad-\frac{s\bar{s}}{2|F^s|^2}\left({\cal D}_{\nu}\frac{\bar{\Omega}^s}{\bar{F}^s}\right)(\g^{\mu\nu\rho}+\g^\mu\eta^{\nu\rho}) P_L\left({\cal D}_{\rho}\frac{\Omega^s}{F^s}\right) -c.c. \Bigg]{\cal D}_\mu\vp\li,
\end{aligned} \end{gather}
where we have put the expression on a simpler and more intuitive form. Note that the spinors and the auxiliary field naturally show up in terms of $P_L\Omega^s/F^s$, and that when a derivative has to act on the spinor for a nonzero expression, it is irrelevant whether this pairing is made explicit or not.

\subsubsection*{Comments}
An expression for $F^\phi$ is easily obtained through $F^\phi=\overline{\bar{F}^\phi}$ and the observation of
\be
\overline{\bar{\Omega}\g^\mu\g^\nu P_L\Omega'}=\bar{\Omega}'P_R\g^\nu\g^\mu\Omega\,. 
\ee
Note that to go from the field definitions of the 4- to the 2-component formalism, as used in this paper, substitutions 
\be
F\rightarrow4F, \qquad
P_L\Omega\rightarrow2\chi, \qquad
P_R\Omega\rightarrow2\bar{\chi},\qquad
{\cal D}\rightarrow 4i{\cal D},
\ee
would be required, due to a different scaling of $\theta$ and the choice of $y^\mu=x^\mu+i\theta\sigma^\mu\bar{\theta}$ in the 2-component notation \cite{notationCam}, different from the 4-component conventions\footnote{Compare (\ref{eq.B},\ref{S},\ref{eq.PhiField}) and \eqref{table} with eq. (14.100) in \cite{Freedman:2012zz}.} of \cite{Freedman:2012zz}. However, further differences may come from the different choices of conventions. In specific, the 2-component relations we have made use of in this article are
\begin{gather}\begin{aligned}
(\theta\psi)&=(\psi\theta)\li,&(\bar{\theta}\bar{\chi})&=(\bar{\chi}\bar{\theta})\li,\\
(\theta\psi)(\theta\psi')&=-\frac{1}{2}\theta^2(\psi\psi')\li,\quad&(\bar{\theta}\bar{\chi})(\bar{\chi}'\bar{\theta})&=-\frac{1}{2}\bar{\theta}^2(\bar{\chi}' \bar{\chi})\li,\\
\theta\sigma^\mu\bar{\theta}\li\theta\sigma^\mu\bar{\theta}&=\frac{1}{2}\theta^2\bar{\theta}^2\eta^{\mu\nu}\li,\\
\overline{\theta\sigma^\mu\bar{\theta}}&=\theta\sigma^\mu\bar{\theta}\li,
\end{aligned}\end{gather}
where substituting either $\psi'=\sigma^\mu\bar{\chi}$, $\psi'=\psi$, $\bar{\chi}'=\psi\sigma^\mu$ or $\bar{\chi}' =\bar{\chi}$ gives further useful relations.

\mciteSetMidEndSepPunct{;\space}{}{\relax}

\bibliography{supergravity}
\bibliographystyle{toinemcite}
\end{document}